\documentclass[prl,floatfix,twocolumn,amsmath,amssymb,superscriptaddress]{revtex4}
\usepackage[pdftex]{graphics}
\usepackage{graphicx}
\usepackage{ulem}

\usepackage{color}

\begin{document}

\title{Large positive magneto-conductivity at microwave frequencies \\ in the compensated topological insulator BiSbTeSe$_2$}

%\author{M.~Bagchi}
\author{Mahasweta~Bagchi}
%\author{L.~Pitz-Paal}
\author{Lea~Pitz-Paal}
%\author{Ch.P.~Grams}
\author{Christoph P.~Grams}
%\author{O. Breunig}
%\author{N. Borgwardt}
\author{Oliver Breunig}
\author{Nick Borgwardt}
\author{Zhiwei Wang}
%\author{Achim Rosch}
%\affiliation{Institute for Theoretical Physics, University of Cologne, Z\"ulpicher Str.\ 77, D-50937 K\"oln, Germany}
\author{Yoichi Ando}
%\author{M. Gr\"uninger}
\author{Markus Gr\"uninger}
%\author{J.~Hemberger}
\author{Joachim~Hemberger}
\email[Corresponding author:~]{hemberger@ph2.uni-koeln.de}
%\affiliation{Physics Institute II, University of Cologne, Z\"ulpicher Str.\ 77, D-50937 K\"oln, Germany}
\affiliation{Physics Institute II, University of Cologne, Z\"ulpicher Str.\ 77, D-50937 Cologne, Germany} %K\"oln

\date{\today}

\begin{abstract}
The bulk electronic properties of compensated topological insulators are strongly affected by the self-organized formation of charge puddles at low temperature, but their response in the microwave frequency range is little studied. We employed broadband impedance spectroscopy up to 5 GHz to address the ac transport properties of well-compensated BiSbTeSe$_2$, where charge puddles are known to form as metallic entities embedded in an insulating host. It turns out that the average puddle size sets the characteristic frequency $\nu_c$ in the GHz range, across which the insulating dc behavior is separated from a metal-like high-frequency response of delocalized carriers within the puddles. 
This $\nu_c$ is found to be controlled by a magnetic field, giving rise to a large positive magneto-conductivity observable only in the GHz range. This curious phenomenon is driven by the Zeeman energy which affects the local band 
filling in the disordered potential landscape to enhance the puddle size. 
\end{abstract}

\maketitle

One of the most prominent features of topological insulators is the spin-momentum locking in the surface states \cite{Ando13}, which is particularly useful for spintronic applications \cite{Pesin12}. To efficiently utilize the spin-momentum-locked surface states in applications, one should get rid of the residual bulk conduction, and this has been achieved by carefully compensating residual donors by residual acceptors through composition tuning in the Bi$_{x}$Sb$_{2-x}$Te$_y$Se$_{3-y}$ compound \cite{Ando13, Taskin11, Ren11}. Although such compensated topological insulators show reasonably high bulk insulation in dc, it has been elucidated that their transport properties are not only determined by the topologically 
protected surface states but also by the presence of charge puddles in the bulk \cite{Skinner12},  
particularly at finite frequencies up to the infrared range \cite{Borgwardt16}.
This is because the Coulomb disorder of the randomly distributed charged donors and acceptors gives rise to potential fluctuations which grow as $\sqrt{R}$ in a volume of size $R^3$ \cite{Skinner12,Shklovskii72}; this inevitably leads to the self-organized formation of charge puddles on a mesoscopic length scale $L$ at low temperatures. On this length scale, portions of a sample contain delocalized charge carriers which give rise to a Drude-like contribution to the optical conductivity $\sigma'(\nu)$ above the Thouless cut-off frequency $2\pi\nu_c$\,=\,$D/L^2$ \cite{Borgwardt16}. This cut-off is given by the time scale needed to diffuse through a puddle with the diffusion constant $D$. Using a simple scaling argument \cite{Skinner12,Shklovskii72}, 
$L \approx 100-500$\,nm is expected on the basis of infrared data \cite{Borgwardt16}. 
This corresponds to $\nu_c$ in the GHz range. 

Since the current-day information technology operates in the GHz range, understanding the microwave response of topological-insulator materials would be of crucial importance for their future applications in spintronics; nevertheless, the role of puddles in the transport properties in the GHz range has not yet been studied. This work is conceived to address this lack of understanding. 

At low temperature, the Drude-like puddle contribution to $\sigma'(\nu)$ and its discrepancy with the dc conductivity have been observed in the well-compensated compound BiSbTeSe$_2$ by Borgwardt \textit{et al.} \cite{Borgwardt16}, who focused on infrared frequencies above the phonon range.
Clear evidence for puddles stems in particular from the characteristic temperature-driven suppression 
of this Drude-like contribution on a temperature scale $k_\text{B} T$\,=\,$E_C$ set by the average Coulomb 
interaction $E_C$ between neighboring dopants \cite{Borgwardt16}. 
This ``evaporation" of puddles is caused by thermally activated carriers which screen the Coulomb disorder. 
In BiSbTeSe$_2$, $E_C/k_\text{B}$\,=\,$30-60$\,K was observed \cite{Borgwardt16,Knispel17}.
Also, puddles can explain the coexistence of electron-type and hole-type carriers in compensated samples \cite{Rischau16} and the small activation energy observed in dc resistivity data \cite{Skinner12}. Intriguingly, puddles are further found to be responsible for the surprising effect of gigantic negative magnetoresistance reported for TlBi$_{0.15}$Sb$_{0.85}$Te$_2$ \cite{Breunig17}.
In this context, novel magnetic-field effects would be expected in the vicinity of the cut-off frequency $\nu_c$. 

In this manuscript, we report on broadband spectroscopic investigations of BiSbTeSe$_2$ in the cut-off frequency range to look for the ac conductivity behavior associated with charge puddles and their response to magnetic fields. 
Our central result is the discovery of a large positive magneto-conductivity at microwave frequencies, which arises due to the change of the puddle size by an external magnetic field. 
In addition, our data provide solid understanding of the role of puddles in the transport properties in the GHz range, %at low temperature, 
providing foundations for future high-frequency applications of topological insulators.

Single crystals of BiSbTeSe$_2$ were grown from a melt of high-purity starting elements Bi, Sb, Te, 
and Se by using a modified Bridgman method in a sealed quartz-glass tube as described elsewhere \cite{Ren11}. 
For the microwave measurements, we used scotch-tape-cleaved flakes with 
length $l\!\approx \!1$\,mm, width $w\!\approx \!1$\,mm, and thickness $t\!\approx \!20$\,$\mu$m; the possible error in the determination of $t$ leads to an uncertainty in the absolute values of about 10\,\%.
The measurements were performed 
in a commercial $^4$He-flow cryo-magnet ({\sc Quantum-Design PPMS}) employing four point geometry for the dc measurements and a home-made, two point coaxial-line inset for the microwave range. 
The complex frequency-dependent conductivity 
$\sigma(\nu)$\,=\,$\sigma'(\nu)+i \sigma''(\nu)$ 
was measured using a coplanar waveguide setup for frequencies up to 5\,GHz evaluating the complex 
scattering coefficients $S_{12}$ and $S_{22}$ 
gained via a vector network analyzer ({\sc ZNB8, Rohde\&Schwarz}). All  measurements were performed with the electric field perpendicular to the crystallographic $c$ axis. The contacts were applied using silver paint. 
The lower boundary of the microwave measurement range is set by the influence of the contact electrodes. 
At the contact interfaces, the formation of Schottky-type depletion layers can lead to Maxwell-Wagner type 
effects \cite{Maxwell91} giving an additional capacitive contribution $C_C$ and
a serial contact resistance $R_C$. 
Together, they form a $RC$ element in series with the intrinsic sample impedance. 
The contacts thus are short-cut for frequencies $2\pi\nu> 1/R_CC_C$ \cite{Lunkenheimer02,Niermann12}. 
This allows us to reliably evaluate the intrinsic sample conductivity for frequencies above 30\,MHz, 
as illustrated in Fig.\ \ref{contactimpedance} of the Supplemental Material \cite{SM}.

The length scale $L$ of puddle formation is connected to a time scale $1/\nu_c$\,=\,$2\pi L^2/D$ 
on which the locally uncompensated carriers behave like free charges. 
For larger times or lengths, carrier motion is hindered by the puddle boundaries.
This leads to insulating behavior for dc or low-frequency transport measurements 
and to a Drude-like metallic conductivity at high frequencies which is cut off below $\nu_c$. 
Assuming that puddles form as soon as the fluctuations of the Coulomb potential are as large as 
$\Delta/2$, where $\Delta$ denotes the band gap,
Shklovskii and coworkers \cite{Skinner12,Shklovskii72} estimated $L$ via a simple scaling argument, 
\begin{equation}
L = \frac{(\Delta/E_C)^2}{8\pi \, N_{\rm def}^{1/3}} \, , 
\label{eq_L}
\end{equation}
where $N_{\rm def}$ is the defect density of acceptors and donors, and the average Coulomb interaction 
between neighboring dopants is given by $E_C$\,=\,$ N_{\rm def}^{1/3} \, e^2/(4 \pi \varepsilon_0 \varepsilon)$ 
with the permittivity $\varepsilon$. 
For BiSbTeSe$_2$, infrared data show $\Delta$\,=\,0.26\,eV and $\varepsilon \!\approx \! 200$ \cite{Borgwardt16}. 
Note that $N_{\rm def}$ and thus also $E_C$ are sample-dependent quantities. 
For two different samples of BiSbTeSe$_2$, $E_C/k_\text{B}$\,=\,$30\!-\!40$\,K and $40\!-\!60$\,K were 
reported, resulting in $N_{\rm def}$\,=\,$0.5-1.1\cdot 10^{20}$\,cm$^{-3}$ and 
$1-4\cdot 10^{20}$\,cm$^{-3}$, respectively \cite{Borgwardt16,Knispel17}. 
This is equivalent to $\Delta/E_C \approx 50-100$ and $L \approx 140-1000$\,nm. 
Using a diffusion constant of $D$\,$\approx $\,2\,cm$^2$/s \cite{Borgwardt16}, 
% and $2\pi\nu_c$\,=\,$D/L^2$, 
we find $\nu_c \lesssim $\,2\,GHz. Note, however, that direct experimental values for $L$ or $\nu_c$ are still lacking.

B\"{o}merich \textit{et al.} \cite{Boemerich17} used a numerical approach combined with a refined 
scaling argument and found 
% much smaller values for the length scale $L$ of puddle formation. 
smaller values for the length scale $L$. 
An experimental lower limit is provided by the length scale $r_s$\,=\,$40-50$\,nm observed for 
\textit{surface} puddles by scanning tunnelling microscopy in BiSbTeSe$_2$ \cite{Knispel17}. 
The Dirac carriers on the surface give rise to enhanced screening, one thus expects $r_s < L$. 
Experimentally, the defect density may not only be sample dependent but it may also vary within 
a given sample. Thus, one has to expect a distribution of puddle sizes and concomitantly a distribution 
of cut-off frequencies, giving rise to a considerably broadened features in the dynamic response.
For BiSbTeSe$_2$, the above discussed limits of $L$, $40$\,nm and $1000$\,nm, span a 
broad frequency range of $0.03$ to $20$\,GHz for the distribution of $\nu_c$.

\begin{figure}[t]
\centerline{\includegraphics[width=0.99\columnwidth,angle=0]{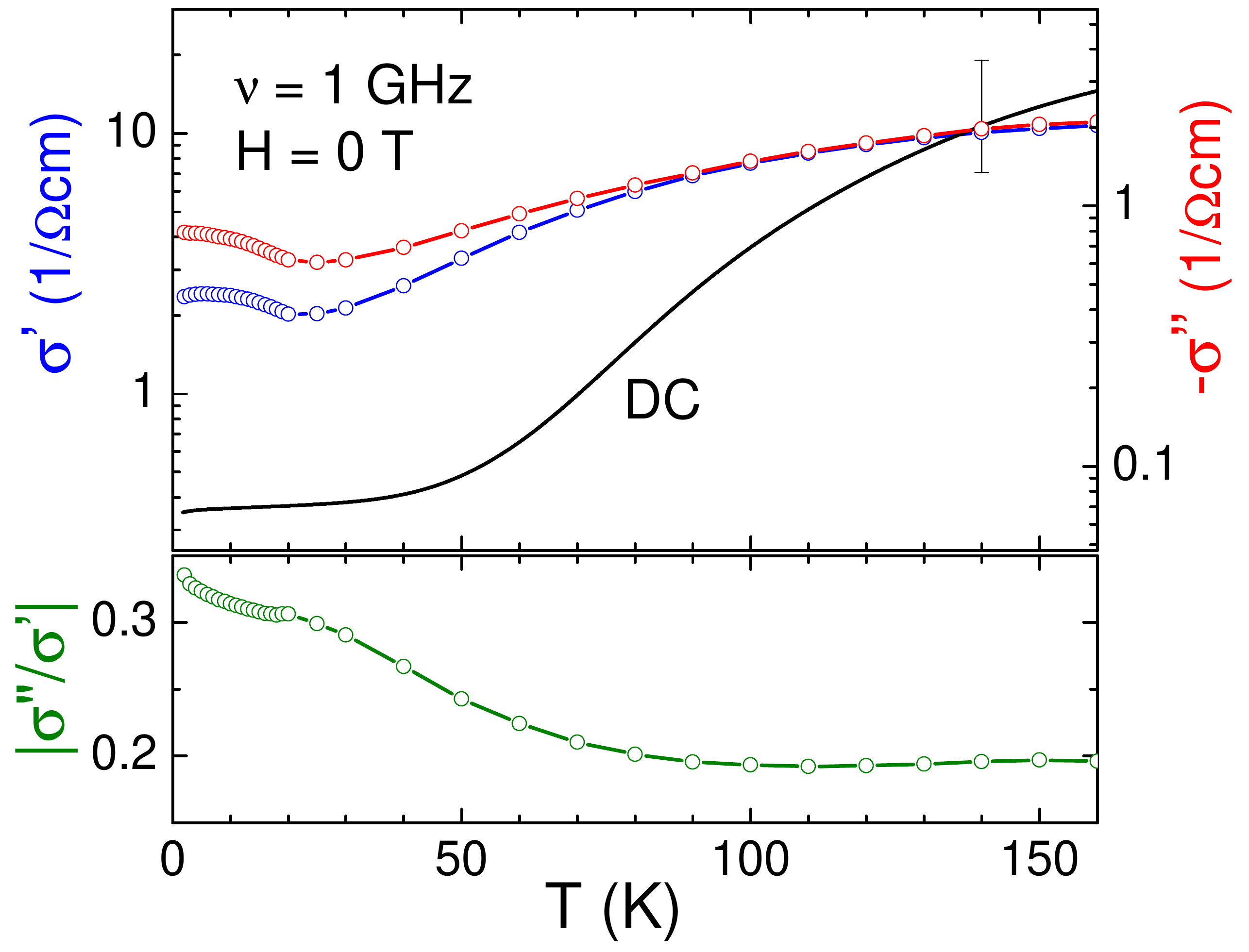}}
\caption{Top: dc and ac (1\,GHz) conductivity $\sigma'$ in compensated BiSbTeSe$_2$. 
%Upon cooling, the reincrease of $\sigma'(1\,$GHz$)$ at low temperature indicates puddle formation. 
%\textcolor{blue}
{The error bar denotes the uncertainty in the absolute values. % of the dc data. 
These vary by about a factor of 3 for samples with the same nominal composition, which partially can be attributed to different defect densities \cite{Ren11}.}
Bottom: 
The ratio $|\sigma''/\sigma'|$ is constant at high temperature, in agreement with variable range hopping. 
For puddle formation, an increase of $|\sigma''/\sigma'|$ at low temperature is expected for frequencies 
close to $\nu_c$.
}
\label{sigma1GHz}
\end{figure}

Clear fingerprints of puddle formation are found in the temperature dependence 
of the conductivity (Fig.~\ref{sigma1GHz}).  
We observe the expected discrepancy between dc and ac conductivities at low temperature, while the high-temperature 
data agree within the experimental error bar. 
At the lowest measured temperatures, the dc conductivity lies more than one decade below the ac data measured 
at $1$\,GHz.
The contribution of puddles is most evident from the reincrease of the ac conductivity below roughly 30\,K, 
which agrees with the energy scale $E_C$ found in the infrared data \cite{Borgwardt16}.  
Note that variable range hopping (VRH) also gives rise to an enhanced ac conductivity, which explains the behavior 
at higher temperatures. 
In BiSbTeSe$_2$, VRH has been shown to describe the dc resistivity in a temperature window 
above about 80\,K \cite{Ren11}.
In the ac conductivity, VRH gives rise to a contribution to 
both $\sigma'(\nu)$ and $\sigma''(\nu)$ 
that scales with a power law 
$\nu^{s}$ 
with $s<1$ \cite{Lunkenheimer03}. 
This scaling behavior results in a fixed ratio $|\sigma''/\sigma'|$ over a wide range in temperature and 
frequency \cite{Jonscher96,Elliott87}. 
In BiSbTeSe$_2$, we find such a constant ratio at 1\,GHz above about 70\,K, see bottom panel of Fig.\ \ref{sigma1GHz}. 
In contrast, the increase of $|\sigma''/\sigma'|$ at lower temperatures can be attributed to the formation 
of polarizable puddles with a finite cut-off frequency $\nu_c$ resulting in a considerable contribution 
to $\sigma''$. 
This additional increase of $\sigma''$ is also evident in the top panel of Fig.\,\ref{sigma1GHz}.

\begin{figure}[t]
\centerline{\includegraphics[width=0.99\columnwidth,angle=0]{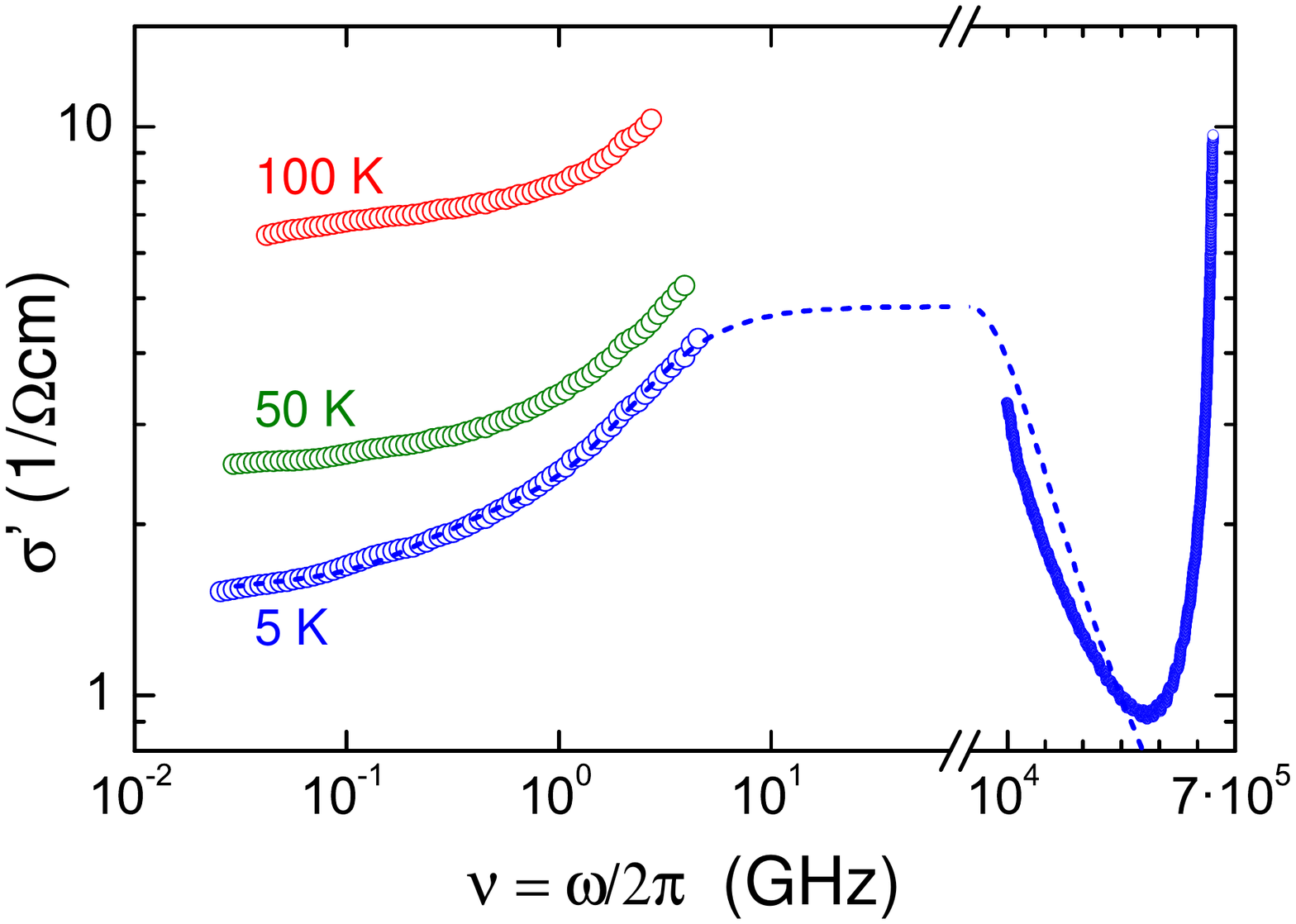}}
\caption{Left part: Broadband microwave conductivity spectra of BiSbTeSe$_2$ on a logarithmic frequency scale for selected temperatures. 
Right part (beyond axis break): Infrared data of a different sample of BiSbTeSe$_2$ \cite{Borgwardt16} on a linear frequency scale. The strong onset of $\sigma'(\nu)$ above about 63\,THz or 0.26\,eV corresponds to the band gap.  
Dashed line: fit to the Drude-like contribution of puddles assuming a distribution of cut-off frequencies as described in the text. The difference in Drude weight between different samples of BiSbTeSe$_2$ can at least partially be attributed to a sample dependence of the defect density, reflecting itself in a variation of the dc resistivity by a factor of about 3. 
}
\label{broadband}
\end{figure}

Figure \ref{broadband} shows broadband microwave spectra of the conductivity $\sigma'(\nu)$ for 
different temperatures. We observe an increase of $\sigma'(\nu)$ with increasing $\nu$ below 5 GHz. Importantly, this increase is more pronounced at 5\,K than at 50\,K or 100\,K, i.e., the frequency dependence is larger at a temperature where puddles have formed. 
The overall much larger value of $\sigma'(\nu)$ at 
higher temperatures and its weaker 
frequency dependence correspond to the metallic Drude contribution of \textit{thermally activated} carriers, as also observed in the infrared range \cite{Borgwardt16}. The low-temperature behavior can be attributed to a smeared-out cut-off of the Drude-like contribution of puddles at frequencies of the order of 1\,GHz. To quantitatively analyse this behavior, we fitted the 5\,K data by considering Drude-Lorentz oscillators, see dashed line in Fig.\ \ref{broadband}: 
In this fitting, the damping $\Gamma$\,=\,$21$\,THz was fixed to the value determined from the infrared data \cite{Borgwardt16}; to take into account the distribution of the puddle size in which the charge carriers are confined, we approximated the situation as a sum %convolution 
of three broad Lorentzians centered at 9.3~GHz, 72~GHz, and 234~GHz (these frequencies are optimized to best reproduce the data). 
For such overdamped oscillators, 
this distribution gives an effective cut-off frequency $\nu_c \! \approx \! \nu_0^2/\Gamma \lesssim 3$\,GHz. Also, this distribution of the eigenfrequencies corresponds to a range of the puddle size $L$ of 0.1 -- 3 $\mu$m, when we use $D$ given in Ref. \cite{Borgwardt16}. 
One can see in Fig. \ref{broadband} that this model reproduces both the low and high frequency behavior of $\sigma'(\nu)$ reasonably well, giving confidence that the overall behavior of $\sigma'(\nu)$ is governed by puddles.

% \subsection{Magnetic field dependence of microwave conductivity}

An external magnetic field influences the size of puddles, as shown recently by Breunig \textit{et al.} \cite{Breunig17} in the nearly bulk-insulating topological insulator TlBi$_{0.15}$Sb$_{0.85}$Te$_{2}$ by dc resistivity measurements. 
Puddles form when the bands touch the chemical potential due to the spatial fluctuations of the Coulomb potential. 
The magnetic field does not change the Coulomb potential but the Zeeman energy shifts the band bottom/top
with respect to the disordered potential landscape and thereby affects the puddle size \cite{Breunig17}. 
The compound TlBi$_{0.15}$Sb$_{0.85}$Te$_{2}$ studied in Ref.\ \cite{Breunig17} is less well compensated 
than BiSbTeSe$_2$ and the low-temperature conductivity is determined by percolating current paths formed 
in the disordered bulk. 
A field-induced change of the puddle size leads to a strong increase of 
percolating puddles and hence a gigantic field-induced reduction of the dc resistance \cite{Breunig17}. 
In contrast, BiSbTeSe$_2$ is well compensated. Puddles do not percolate, the bulk resistivity at low temperatures 
is much higher, and the influence of a magnetic field on the dc conductivity is expected to be 
much weaker as long as the percolation limit is not reached. Close to the cut-off frequency, however, 
a magnetic-field-induced change of the puddle size may have considerable impact.

\begin{figure}[t]
\centerline{\includegraphics[width=0.99\columnwidth]{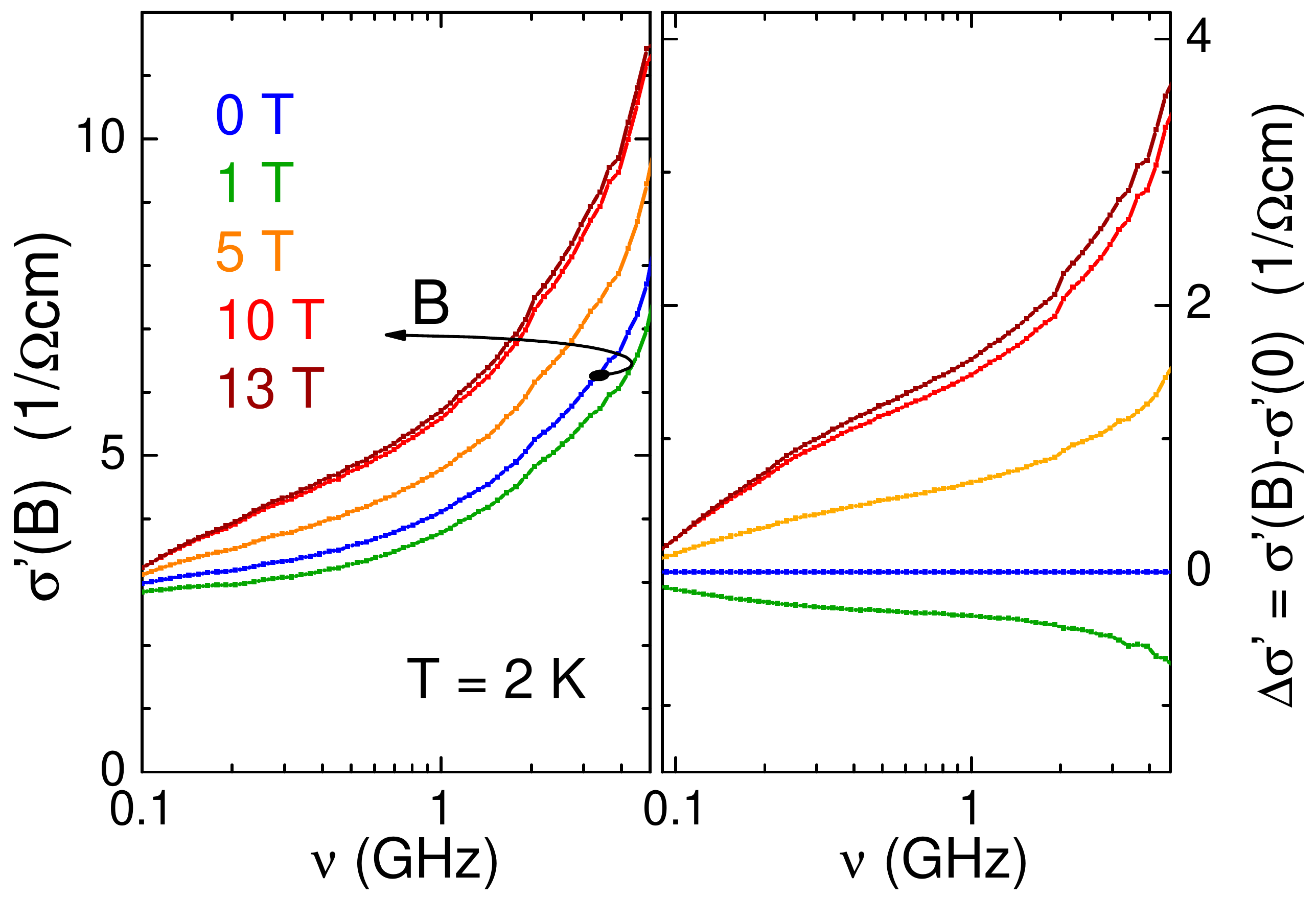}}
\caption{
Left: Conductivity spectra measured at $T$\,=\,$2$\,K for various values of the external magnetic field up to $13$\,T. 
Right: Magnetic field dependent contribution to the conductivity spectra as gained by subtraction of the zero field data. % $\Delta\sigma'(\nu)$.
}
\label{cutoff-shift}
\end{figure}

Figure \ref{cutoff-shift} shows $\sigma'(\nu)$ at 2\,K for different values of the external magnetic field. One can see that the behavior at 1 T is clearly different from that at higher fields; namely, the conductivity in 1 T is lower than in zero field and the difference increases with increasing frequency, as is evident in the right panel. At such low fields, the ac transport is dominated by weak anti-localization (WAL) characteristic for disordered systems with strong spin-orbit coupling \cite{Altshuler81,Ando13,Hikami80}.
The suppression of WAL in small magnetic fields leads to localization of carriers and therefore to a reduced conductivity. 
In contrast, in higher fields the conductivity increases strongly, in particular at higher frequencies. 
This can be attributed to a field-induced shift of the cut-off frequency $\nu_c$ that is triggered by a field-induced change of the puddle size. This shift is directly visualized in the left panel of Fig.\,\ref{cutoff-shift} via the shift of the rise in $\sigma'(\nu)$.

\begin{figure}[tb]
\centerline{\includegraphics[width=\columnwidth]{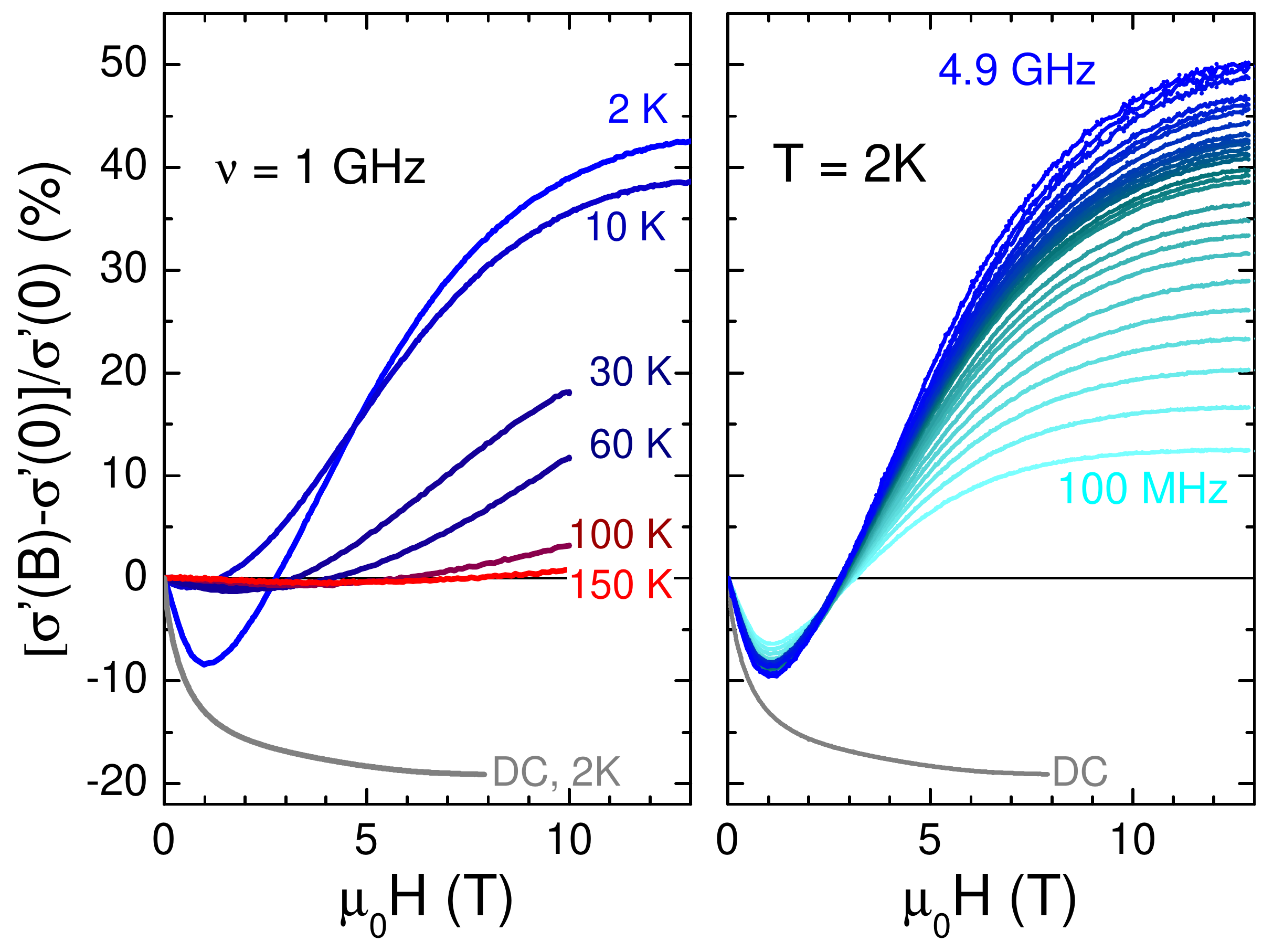}}
\caption{Left: Normalized magneto-conductivity measured at 1\,GHz for various temperatures from 2\,K to 150\,K.  
Right:	Normalized magneto-conductivity measured at $T$\,=\,$2$\,K for different frequencies in equidistant logarithmic spacing between 100\,MHz and 4.9\,GHz.\@ 
		The dc conductivity at 2~K (gray) is displayed for comparison. 
		}
\label{magneto}
\end{figure}

At high fields, the suppression of weak anti-localization still applies but it is overruled by the strong positive ac magneto-conductivity. 
This is evident in the left panel of Fig.\,\ref{magneto},
which depicts the normalized magneto-conductivity 
%${\sigma'(B)/\sigma'(0)}-{1}$ 
${[\sigma'(B)-\sigma'(0)]/\sigma'(0)}$ 
at 1 GHz as a function of the external magnetic field. The ``dip'' below 3 T is due to the WAL effect discussed above, but a pronounced positive magneto-conductivity sets in at higher fields. As shown in the right panel of Fig. 4, the high-field magneto-conductivity strongly increases with frequency, reaching 50\% in 13 T at 4.9\,GHz.
The dc data shown in Fig. 4 for comparison is governed by the WAL all the way up to high fields and only presents negative magneto-conductivity approaching $-20$\% at high magnetic field. 
This crossover in the high-field magneto-conductance from negative to positive upon changing from dc to the microwave range shows that it is not a field-induced percolation which dominates transport properties in compensated BiSbTeSe$_2$. Rather, %it seems that 
the magnetic field mainly affects the average size of still isolated puddles. The corresponding shift of the cut-off frequency $\nu_c$ naturally explains the large magneto-conductivity at frequencies in the vicinity of $\nu_c$ in the GHz range.

In the left panel of Fig.\,\ref{magneto}, one can also see how the magneto-conductivity behavior at 1\,GHz changes with temperature. 
The pronounced positive microwave magneto-conductivity is limited to the low-temperature range where charge puddles are present. At higher temperatures ($\gtrsim$ 30 K) where puddles have evaporated, the magneto-conductivity is drastically suppressed.

% \subsection{Summary}

In conclusion, we have demonstrated that the ac transport properties of the compensated topological insulator 
BiSbTeSe$_2$ at low temperatures are determined by the presence of charge puddles, which gives rise to an unusual positive magneto-conductivity observable only in the GHz range. 
The random nature of the Coulomb-potential fluctuations causes a distribution of the puddle size, which
leads to a broad cut-off regime in the microwave range separating the insulating dc behavior from the 
metal-like ac response; our analysis shows that the overall ac conductivity behavior points to the puddle-size distribution of 0.1 -- 3 $\mu$m. 
The remarkable effect of the magnetic field to enlarge the puddle size, previously suggested in Ref. \cite{Breunig17} for an imperfectly compensated topological insulator, is now found to cause a large positive magneto-conductivity even in a well-compensated system near the characteristic cut-off frequency. Hence, tailoring the defect density and the corresponding disorder-induced microstructure will be of crucial importance to any potential applications of bulk topological insulators in the technologically important GHz range.

{\bf Acknowledgements}

The authors thank A. Rosch for valuable discussions. This work was funded by the Deutsche Forschungsgemeinschaft (DFG, German Research Foundation) 
-- Project number 277146847 -- CRC 1238 (projects A04 and B02).

%%%%%%%%%%%%%%%%%%%%%%%%%%%%%%%%%%%%%%%%%%%%%%%%%%%%%%%%%%%%%%%%%%%%%%

\clearpage
% \newpage

\renewcommand{\theequation}{S\arabic{equation}}
\renewcommand{\thefigure}{S\arabic{figure}}
\renewcommand{\thetable}{S\,\Roman{table}}

\setcounter{figure}{0}

\begin{flushleft} 
{\Large {\bf Supplemental Material}}
\end{flushleft}

\begin{flushleft}
{\bf Effect of Electrical Contacts}
\end{flushleft}

Schottky-type depletion layers tend to form at the electrical contact formed at interfaces between sample and silver electrodes. The depletion layers lead to Maxwell-Wagner type contributions to the complex impedance $Z^*(\omega)$ \cite{Maxwell91}. These can be modelled employing an equivalent circuit containing an additional capacitive contribution $C_C$ and a contact resistance $R_C$. Together, they form an $RC$ element in series with the intrinsic sample impedance. The resulting impedance thus can be described as
%\[
%Z^* = \frac{R_c}{1 + \omega^2R_c^2C_c^2} 
%		+ \frac{R_s}{1 + \omega^2R_s^2C_s^2}
%	 - i\left(
%		\frac{R_C \omega R_c C_c}{1 + \omega^2R_c^2C_c^2} 
%		+ \frac{R_s \omega R_s C_s}{1 + \omega^2R_s^2C_s^2}
%		\right)
%\]
\[
Z^* = \frac{R_c}{1 + \omega^2\tau_c^2} 
+ \frac{R_s}{1 + \omega^2\tau_s^2}
 - i\left(
\frac{R_C \omega \tau_c}{1 + \omega^2\tau_c^2} 
+ \frac{R_s \omega \tau_s}{1 + \omega^2\tau_s^2}
\right)
\]
with the relaxation times $\tau_c=R_cC_c$ and $\tau_s=R_sC_s$ for contacts and sample, respectively. Usually, the thin contact depletion layers lead to $\tau_c \gg \tau_s$ and the contacts thus are short-cut for frequencies $2\pi\nu > 1/\tau_c$ \cite{Lunkenheimer02,Niermann12}. 
In addition one can expand the accessible frequency range by subtracting the contact contribution, which dominates the low frequency range and therefore can be fitted for $2\pi\nu<1/\tau_c$. The remaining impedance can be evaluated via the geometry factor $A/d$ with respect to the complex conductivity 
$\sigma' = A/d \cdot Z'/Z^2$ and $\sigma'' = A/d \cdot Z''/Z^2 $.

\begin{figure}[h]
\centerline{\includegraphics[width=0.9\columnwidth]{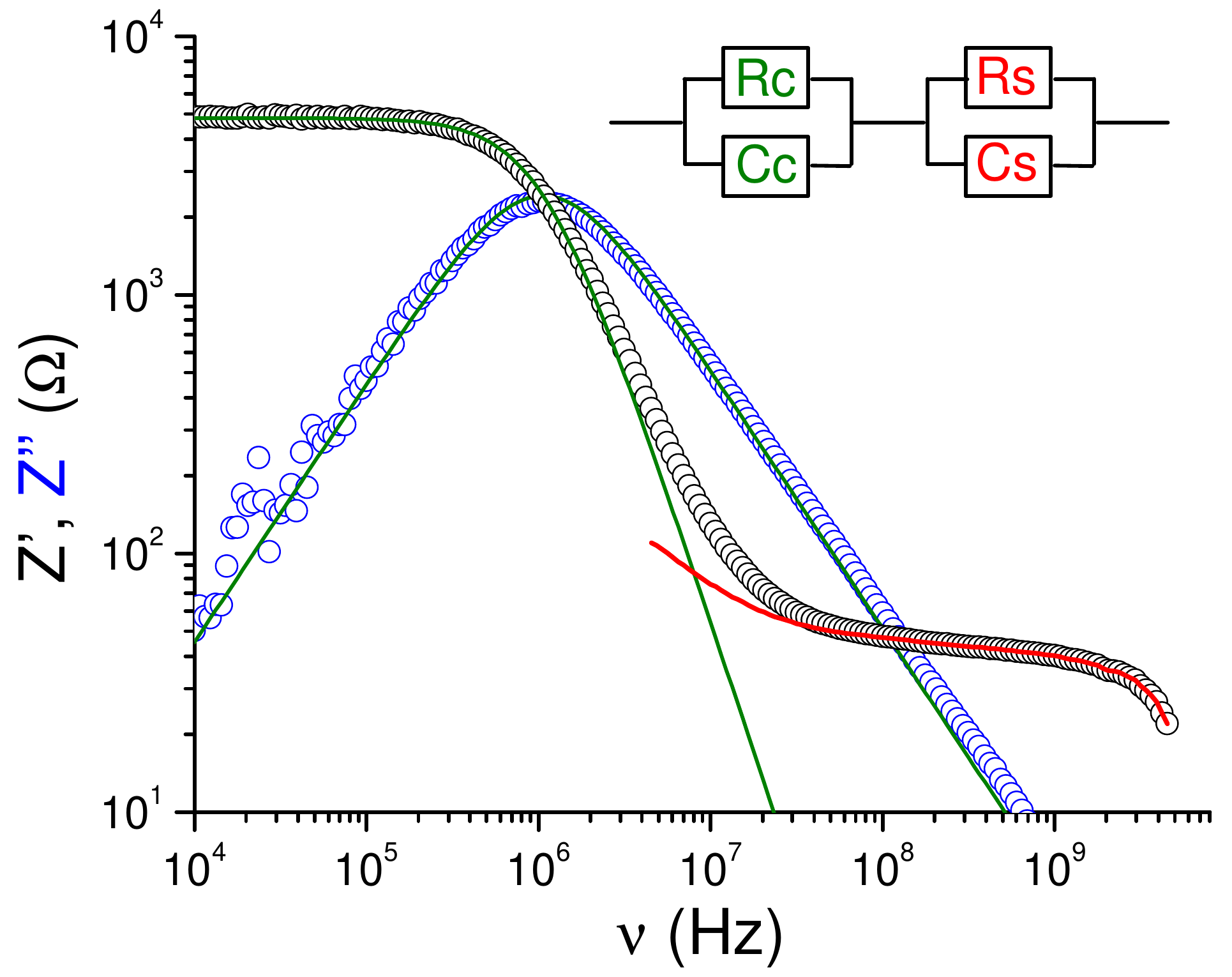}}
\caption{Frequency dependence of the measured complex impedance $Z^*$ at 300\,K (circles), 
which is dominated by the contacts at low frequencies and by the sample impedance at high frequencies.  
The contribution of the contacts is depicted by the solid green line, assuming an $RC$ element with constant values for $R_C$ and $C_C$. 
The solid red line represents the intrinsic sample contribution given by $R_S(\nu)$ and $C_S(\nu)$.
The sketch in the inset shows the corresponding equivalent circuit.
}
\label{contactimpedance}
\end{figure}

\end{document}